\documentclass[12pt a4paper]{revtex4}
\usepackage{setspace}
\usepackage{graphicx}
\usepackage{amsmath}
\usepackage{amsfonts}
\usepackage{amssymb}
\begin{document}

\title{Weak polyelectrolytes in the presence of counterion condensation with ions of variable size and polarizability}
\author{Prasanta Kundu and Arti Dua}
\affiliation{Department of Chemistry, Indian Institute of Technology, Madras, Chennai-600036, India}

\date{\today}

\begin{abstract}
\noindent
Light scattering and viscometric measurements on weak polyelectrolytes show two important aspects of counterion condensation, namely,  non-monotonic variation in the polyelectrolyte size with the increase in the electrostatic strength, and,  monovalent counterion selectivity in determining the nature of collapse transition  at high electrostatic strengths. Here, we present a self-consistent variational theory for weak polyelectrolytes which includes the effects of the polarizability of monovalent counterions. Our theory reproduces several experimental findings including non-monotonic conformational size with the variation in the electrostatic strength and a shift from a continuous to a discontinuous collapse transition with the increase in the dipole strength of condensed ions. At low dipole strength and high electrostatic strength, our theory predicts a series of solvent quality driven size transitions spanning the re-entrant poor, theta and good solvent regimes. At high dipole strength, the size remains that of a compact globule  independent of solvent quality. The dipole strength of the ion-pair formed due to counterion condensation, which depends on the size and polarizability of the monovalent counterions, is found to be an important molecular parameter in determining the nature of collapse transition, and the size of the collapsed state at high electrostatic strength.
\end{abstract}
\maketitle

\section{Introduction}
Weak polyelectrolytes are polymers with ionizable functional groups that partially dissociate in polar solvents leaving ions of one sign bound to the polymer chain and oppositely charged counterions in solution \cite{Barrat,Hara}. In contrast to strong polyelectrolytes which dissociate completely in the entire pH range accessible experimentally, weak polyelectrolytes carry weak acidic or basic functional groups whose degree of ionization can be controlled by the change in the solution pH \cite{Zito,Raphael,Uyaver}. The conformational behaviour of weak polyelectrolytes can thus be finely tuned by changes in the solution pH, ionic strength, solvent quality and temperature rendering them with properties which find applications in many areas including gene delivery \cite{gene}, regulation of DNA transcription, DNA electrophoresis, water ultrafiltration \cite{ultrafiltration} and purification \cite{purification}. 

Conformational behaviour of partially ionized poly(acrylic acid) (PAA) in a dilute solution at room temperature has been observed using viscometric titrations against strong bases like CH$_3$OLi and CH$_3$ONa \cite{Klooster}. Among other things, these measurements show non-monotonic variation in the reduced viscosity with the increase in the degree of ionization. While the reduced viscosity is directly proportional to the polymer size through the Flory-Fox equation and provides an indirect measure of the polyelectrolyte size, the increase in the degree of ionization of PAA increases the fraction of charges along the chain, thereby increasing the strength of the electrostatic interaction. The increase in the reduced viscosity at low degree of ionization is due to electrostatic repulsion between like charged monomers along the polyelectrolyte backbone resulting in polyelectrolyte chain swelling. The decrease in the reduced viscosity at high degree of ionization is more intriguing, and believed to be due to attractive interaction between ion-pairs (dipoles) formed between partially charged polyelectrolyte and oppositely charged counterions. This effect due to counterion condensation results in polyelectrolyte chain collapse at high degree of ionization. Interestingly, the nature of the collapse transition depends on the size and polarizability of the monovalent counterions, exhibiting continuous reduction in the polyelectrolyte size  for smaller and less polarizable Li${^+}$ as compared to discontinuous collapse transition for larger and more polarizable Na$^+$ \cite{Klooster}. In the former case, the reduced viscosity at high degree of ionization is close to its initial value at vanishingly small degree of ionization, reminiscent of the re-entrant initial size at high degree of ionization. In the latter case, the reduced viscosity at high degree of ionization is much smaller than its initial value at vanishingly small degree of ionization, indicating the polyelectrolyte chain collapse. Moreover, the chain collapse in the presence of Na$^+$ as counterions occurs for much lower electrostatic strength compared to Li$^+$ as counterions. Light scattering and osmotic pressure measurements on PAA observe similar trends \cite{Klooster}. 

For an infinitely long charged rod, the phenomenon of counterion condensation is commonly described using Manning's theory \cite{Manning}. The latter shows reduction in the fraction of charges along the polyelectrolyte backbone with the increase in the dimensionless electrostatic strength, $l_B/b$, where $l_B$ is the Bjerrum length and $b$ is the effective distance between monomers.  For a charged polymer, the theory of counterion condensation due to Manning is only applicable for stiff rod-like polyelectrolytes, where the persistence length of the chain is much larger than the chain length. The presence of the long-range electrostatic interactions in combination with the polymer elasticity make the phenomenon of counterion condensation in flexible polyelectrolytes much more complex than the Manning counterion condensation, and has been the focus of several analytical theories and simulations on strong polyelectrolytes \cite{Pincus,Muthukumar,Brilliantov,Limbach,Winkler,Sunil,Anup} . These studies predict non-monotonic variation of the polyelectrolyte size as a function of the electrostatic strength. Simulations also observe the formation of ion-pair due to counter-ion condensation at high electrostatic strength \cite{Limbach,Winkler,Sunil,Anup}. However, there is discrepancy in the nature of collapse transition at large electrostatic strength, which is found to be either discontinuous or continuous.  In recent simulations of a polyelectrolyte  chain in a poor solvent, for instance, the reduction in the polyelectrolyte size at high electrostatic strength is found to be continuous \cite{Limbach,Sunil}. The reduction in the chain size occurs via a sausage-like phase, and results in a re-entrant poor solvent size at high electrostatic strength \cite{Limbach,Sunil}. Another recent simulation, on the other hand, observes a discontinuous chain collapse without the formation of a sausage-like phase,  and argues that a continuous collapse transition  is an artifact  of finite size effects, arising from considering a short polyelectrolyte chain with very low degree of polymerization, $N$ \cite{Anup}. While it is not clear if $N= 199$ and $N=120$ considered in references \cite{Limbach}  and \cite{Sunil} respectively represent chains with very low degree of polymerization, another theoretical study observing a discontinuous collapse transition suggests that the nature of collapse transition can shift from  discontinuous to continuous by varying certain molecular parameters \cite{Pincus}. Although which molecular parameter can bring about such a shift is not clear, viscometric  measurements on PAA with molar mass $5.5 \times 10^5$ g mol$^{-1}$ ($N \approx 7640$) provide evidence that even for chains with very high degree of polymerization the increase in the size and polarizability of the condensed counterions, corresponding to increase in dipole strength of the ion-pairs, shows a shift from a continuous to a discontinuous collapse transition \cite{Klooster}. These studies indicate that there are currently discrepancies between theory, simulation and experiment which need to be resolved.

In this work, we present a self-consistent variational theory which includes the effects of the dipole strength of the ion-pairs, formed between condensed counterions and oppositely charged monomers at high electrostatic strengths, in determining the nature of collapse transition in weak polyelectrolytes. The theory subsumes all previous results and unifies them in a single framework.  In particular, a continuous reduction in the polyelectrolyte size with a re-entrant poor, theta or good solvent size at high electrostatic strength and a discontinuous collapse transition to compact globular state independent of the solvent quality observed independently in references \cite{Limbach,Sunil} and  \cite{Anup} respectively are both captured in the present theory by just increasing the dipole strength of the ion pairs, in qualitative  agreement with experiments \cite{Klooster}.  The dipole strength of the ion-pairs, which depends of the size and polarizability of the condensed counterions at high electrostatic strength, emerges as an important molecular parameter in influencing the nature of collapse transition and the size of the collapsed state at high electrostatic strength. 


The paper has been organized as follows. In Section II, a self-consistent variational theory  to determine the size and fraction of charges of a weak  polyelectrolyte is presented in the presence of counterion condensation.  In Section III,  the numerical solution of the coupled non-linear equations for the size and the fraction of charges is presented. The effects of  counterion condensation in determining the fraction of charges along the polyelectrolyte backbone and conformational transitions of a weak polyelectrolyte as a function of the solution pH, solvent quality, electrostatic interaction strength $l_B$, the degree of polymerization for different values of the dipole strength are also discussed in Section III.  Section IV provides a brief summary of the results. The details of the calculations are presented in Appendices A, B and C.


\section{Theory}
The starting point of our calculation is the following expression for the Hamiltonian, which describes the conformation of a weak and flexible polyelectrolyte \cite{Muthukumar,Prasanta,Singh,Dua} 

\begin{equation}
{\cal H} = {\cal H}_0 + {\cal H}_{2} + {\cal H}_3 + {\cal H}_c
\end{equation}

and  
\begin{eqnarray}\label{Ham0}
\beta{\cal H} &=& \frac{3}{2\,b^2} \int _{0}^{N} dn \left(\frac{\partial{\bf r}_n}{\partial n}\right)^2 +
\frac{v_2}{2} \int_{0}^{N} dn \int_{0}^{N} dm \,\delta\,[{\bf r}_n - {\bf r}_m] \nonumber\\
& & + \frac{w}{6} \int_{0}^{N} dn \int_{0}^{N} dm \int_{0}^{N} dl \,\delta\,[{\bf r}_n - {\bf r}_m] \,\delta\,[{\bf r}_m - {\bf r}_l] \nonumber\\
& &  + \frac{f^2\, l_B}{2} \int_{0}^{N} dn \int_{0}^{N} dm \, \frac{e^{-\kappa |{\bf r}_n - {\bf r}_m|}}{|{\bf r}_n - {\bf r}_m|} , 
\end{eqnarray}\\
where the polymer conformation is described by the position vector ${\bf r}_n$ of the $n$th monomer; $b$ is the effective bond length or Kuhn length and $N$ is the effective number of monomers. The first term describes the entropic elasticity of the chain; the second and the third terms represent the short range two-body  and three-body interactions respectively;  $v_2$ and $w$  represent the strengths of the two-body and three-body interactions respectively. The fourth term represents the long range screened Coulombic interactions, where $f$ is the fraction of charges on the chain backbone. $l_B = e^2/4\pi \epsilon k_B T$ is the Bjerrum length, where $e$ is the elementary charge and $\epsilon$ is the dielectric constant of the solvent. The Bjerrum length defines the length scale at which the electrostatic energy is of the order of $k_B T$. In the absence of salt, $\kappa$ represents the inverse Debye screening length due to the presence of the oppositely charged counterions. Since the fraction of charges on the chain backbone depends on the pH of the solution, the expression for the inverse screening length is given by $\kappa = (4 \pi\,\l_B ( f \rho + 2 \,(1-f)\rho))^{1/2}$, where $\rho$ is the density of monomers.  The first term  in the last expression is due to free counterions and the second term is due to the hydroxide (or methoxide) ions in hydrolysis of the polyelectrolyte backbone \cite{Yethiraj}.

When counterions condense on the polyelectrolyte chain, they form ion-pairs (dipoles) with the oppositely charged monomers, with dipole moment $P = e d$, where $d$ is the distance between the charge on the monomer and the oppositely charged counterion. The dipolar interaction, which is short ranged, modifies the strength of the two-body interaction \cite{Pincus}. From the Mayer $f$-function, $f(r) = e^{-\beta U(r)} - 1 $, the strength of the two-body interaction term is given by \cite{Rubinstein}
\begin{equation}\label{ex-vol}
v_2 = -4\pi \int_0^{\infty} d{r}\, r^2 f(r) = 4\pi\int_0^{\infty} d{r}\, r^2 (1 - e^{-\beta U(r)}) .
\end{equation}
Assuming that the condensed counterions only slightly perturb the repulsive and attractive interactions, the two-body interaction energy is given by $U(r) = U_{rep}(r) + U_{att}(r) + U_d(r)$ \cite{Pincus}, where the first two terms represent the non-electric part of the interaction energies comprising of  the hard core repulsion term and the weak attractive potential term respectively. The last term is due to the dipole interaction, $U_d(r)/ k_B T = - 4 \pi (1-f)^2\, l_B^2 \,d^4/3\, r^6$, at distance  $r$ between the two dipoles \cite{Dipole}. For $r < b$,\, $U_{rep}/ k_B T \gg 1$ and the only contribution to the above integral comes from the repulsive part of the potential. For $r > b$, on the other hand,  the attractive part dominates and the only contribution to the above integral comes from the attractive part of the potential \cite{Rubinstein}. The calculation of the integral in Eq. (\ref{ex-vol}), thus, yields 
\begin{equation}
v_2 \approx v -\frac{16\,\pi ^2\,(1-f)^2\,l_B^2\,d^4}{9 b^3}.
\end{equation}
The details of the calculation are given in Appendix A. In the above expression, $v$ is the strength of the excluded volume interaction in the absence of counterion condensation, the sign of which depends on the solvent quality. The sign of $v$ is positive in good solvents and negative in poor solvents. In theta solvents $v = 0$. 

It is to be noted that in writing the expression for $U(r)$ in Eq. (\ref{ex-vol}), we have ignored the monopole-dipole interaction. Although both monopole-dipole and dipole-dipole interactions are short-ranged attractive interactions and scale as $-f(1-f)l_B^2 d^2/r^4$ and $-(1-f)^2l_B^2 d^4/r^6$ respectively and contribute at high $l_B$, the contribution from the dipole-monopole interaction is much smaller than the the dipole-dipole interaction interaction  at small $r$ and high $l_B$ for two reasons. First, because dipole-dipole interaction is of much shorter range $1/r^6$ and is more relevant at smaller $r$ and high $l_B$; second, the dipole-dipole strength, $d^4 (1-f)^2 $, is higher than the dipole monopole strength, $d^2 f (1-f)$ at large $l_B$. This is because as $f$ get renormalized  and become smaller at high $l_B$ due to counterion condensation, the contribution from  monopole-dipole interaction becomes smaller as it depends on $f(1-f)$ compared to $(1-f)^2$ dependence of the dipole-dipole interaction.  Recent simulations have shown that an increase in $l_B/b$ from one to slightly greater than one leads to 80\% condensation of oppositely charge ions to form dipoles \cite{Sunil,Muthukumar2002}. This indicates the relevance of dipole-dipole interactions, which  for this reason is usually ignored in comparison to monopole-dipole interactions \cite{Cherstvy2010}.

Below, we use the uniform expansion method of Edwards and Singh \cite{Singh,Doi}, to determine the size of a polyelectrolyte chain in good, poor and theta solvents. The uniform expansion method is a self-consistent variational approach, originally used to determine the size of a neutral polymer in good solvents \cite{Singh, Doi}. It was later extended to study polyelectrolytes and polyampholytes in good \cite{Muthukumar87} and  poor solvents \cite{Prasanta,Dua,Thirumalai}.  In the absence of any interaction term, the probability distribution of the chain end-to-end distance is Gaussian, and yields the mean square end-to-end distance as $\left< {\bf R}_0^2 \right> = N b^2$.  The uniform expansion method \cite{Singh} defines a new step length, $b_1 \gg b$, such that the mean square end-to-end distance of the chain in the presence of the interaction terms   is $\left< {\bf R}^2 \right> = N b_1^2$. For this, it is required that the original Hamiltonian ${\cal H}_{0} = \frac{3}{2b^2} \int _{0}^{N} dn {\dot{\bf r}}_n^2$ is replaced with the reference Hamiltonian ${\cal H}_{1} = \frac{3}{2b_1^2} \int _{0}^{N} dn {\dot {\bf r}}_n^2$. This method, which is routinely used to calculate the polymer size in the presence of the interaction terms, is outlined in the book by Doi and Edwards for a neutral polymer in a good solvent \cite{Doi}.  Using this method, the expression for the mean square end-to-end distance of a polyelectrolyte in different solvents in the presence counterion condensation  is given by
\begin{equation}
\label{Ham}
\left< {\bf R}^2 \right> = 
\frac{\int {\cal D}[R_n]\,{\bf R}^2 \,\exp[-({\cal H}_1+ 
({\cal H}_0-{\cal H}_1)+{\cal H}_{2}+{\cal H}_3 + {\cal H}_c)]}{\int {\cal D}[R_n]\,
\exp[-({\cal H}_{1}+ ({\cal H}_{0}-{\cal H}_{1})+{\cal H}_{2}+{\cal H}_3 +{\cal H}_c)]}.
\end{equation} \\[1ex]
 
The last equation is the definition of $\left<{\bf R}^2\right>$, but by adding and subtracting  ${\cal H}_{1}$ from ${\cal H}_{0}$, $\left< {\bf R}^2 \right>$ can be expanded in a perturbative series about the reference
Hamiltonian ${\cal H}_{1}$ to yield \cite{Prasanta}. 
\begin{equation}
\label{perturb}
\left< {\bf R}^2 \right> \left< {\cal H}_{0}-{\cal H}_{1}+{\cal H}_2+ {\cal H}_3 + {\cal H}_c\right> - \left< {\bf R}^2 \,({\cal H}_{0}-{\cal H}_{1}+{\cal H}_2+ {\cal H}_3  + {\cal H}_c) \right> = 0. 
\end{equation}
Eqn.(\ref{perturb}) can be written as
\begin{eqnarray}
\label{Hamiltonian}
& &\left[\left<{\bf R}^2 \right> \left< {\cal H}_0 - {\cal H}_1 \right> - \left<{\bf R}^2\left({\cal H}_0-{\cal H}_1\right)\right> \right] + 
\left[\left<{\bf R}^2\right> \left<{\cal H}_2\right> - \left<{\bf R}^2\left({\cal H}_2\right)\right> \right] + \left[\left<{\bf R}^2\right> \left<{\cal H}_3\right> - \left<{\bf R}^2\left({\cal H}_3\right)\right> \right] \nonumber \\
& & \hspace{9.5cm}+ \left[\left<{\bf R}^2\right> \left<{\cal H}_c\right>  - \left<{\bf R}^2 \left({\cal H}_c\right)\right> \right] = 0.
\end{eqnarray}
We solve each of these terms, the details of which are provided in Appendix B. The resulting equation is given by
\begin{equation}
\label{eqn1}
1 - \alpha^2 + \frac{\sqrt{6}\, v\, N^{1/2}}{\pi^{3/2}\,\alpha^3\, b^3}  + \frac{1}{\alpha^6} + \frac{ f^2 \,l_B\, N^{3/2}\, \Lambda(\kappa_0)}{6\, \pi^2 \,b\,\alpha} - \frac{ 16\,\sqrt{6}\,\pi^2 \,(1-f)^2\, l_B^2 \,d^4\, N^{1/2}}{9\,\pi^{3/2}\,b^6\,\alpha^3}   = 0,
\end{equation}
with
$$
\label{salt}
\Lambda(\kappa_0) = \int_{0}^{\infty} dk_0\int_{0}^{1} dy \int_{0}^{1} dx\, \frac{(1 - x)^2 \,y^3 \,k_0^4\, e^{-k_0^2\, y \,(1 - x)}}{(k_0^2 + \kappa_0^2)} 
$$
where $\alpha$ is a measure of the dimensionless average size given by $\alpha = \sqrt{\frac{\left<{\bf R}^2\right>}{\left<{\bf R}_0^2\right>}} = \frac{b1}{b}$, where $\left<{\bf R}^2\right>$ is the average size of a polyelectrolyte in the presence of the interaction terms and $\left<{\bf R}_0^2\right>$ is the average size of an ideal chain in the absence of the interaction terms; $\kappa_0 = \kappa\,N^{1/2}\,b_1$ is the dimensionless inverse screening length; $k_0 = k\, N^{1/2} \,b_1$, $x = m/n$ and $y = n/N$ are the dimensionless integration variables. The effective free energy of a polyelectrolyte chain can be obtained from Eq. (\ref{eqn1}) by using the method of thermodynamic integration \cite{Prasanta,Allen}
\begin{equation}
\label{free-energy1}
\beta F_{chain}({\alpha}, f) = -\ln\alpha+\frac{\alpha ^2}{2}+\left(\frac{2}{3\,\pi}\right)^{1/2}\frac{v\,N^{1/2}}{\pi\,b^3\,\alpha ^3}+\frac{1}{6\,\alpha ^6}+\frac{f^2 \,l_B \,N^{3/2}\,\Lambda(\kappa_0)}{6\,\pi ^2\, b\,\alpha} -\frac{16\,\sqrt{6\,\pi}\,(1-f)^2\,l_B^2\,d^4\,N^{1/2}}{27\,b^6\,\alpha ^3}.
\end{equation}
In the method of thermodynamic integration, the change in the free energy can be obtained from the integration of the free energy gradient, $\left(\frac{\partial F_{chain} (\alpha', f)}{\partial {\bf \alpha'}}\right)_f$, the minimization of which yields the equilibrium path connecting the two states. Thus, Eq. (\ref{free-energy1}) is obtained by using  $F_{chain}(\alpha , f) = \beta F_{chain} (\alpha_0, f) + \int_{\alpha_0}^{\alpha}  \left(\frac{\partial F (\alpha', f)}{\partial {\bf \alpha'}}\right)_f d\alpha^{\prime}$, where  $\alpha_0 $ is an arbitrary reference state. The first two terms in Eq. (\ref{free-energy1}) are due to the entropic elasticity of the chain, the third term is due to the excluded volume interaction, which accounts for the solvent quality. The fourth term is due to the three-body repulsion interaction. The fifth term is due to the screened electrostatic interaction between charged monomers. The sixth term accounts for the dipolar interaction between ion-pairs formed between the charged monomers on the polyelectrolyte chain and the oppositely charged counterions. The electrostatic contribution to the free energy of the polyelectrolyte chain is, thus, given by 
\begin{equation}
\beta F_{el} = \frac{f^2 \,l_B \,N^{3/2}\,\Lambda(\kappa_0)}{6\,\pi ^2\, b\,\alpha} - \frac{16\,\sqrt{6\,\pi}\,(1-f)^2\,l_B^2\,d^4\,N^{1/2}}{27\,b^6\,\alpha ^3}.
\end{equation}
In the presence of counterion condensation, the fraction of charged monomers on the chain, given by $f$ in Eq. (\ref{free-energy1}), are not fixed but depends on the pH of the solution. In a mean-field approximation the grand-canonical free energy for condensed and uncondensed counterions (CI) is given by
\begin{equation}\label{free-energy-CI}
\beta F_{CI} (\alpha, f) =  N[ f \log_{10} f + (1-f) \log_{10} (1-f) + f \mu] + \beta F_{el} + \beta F_{fl},
\end{equation}
where $\mu = pH - pK_0$ and $K_0$ is the intrinsic dissociation constant \cite{Barrat,Zito,Raphael,Uyaver}. In the above equation, the first two terms are due to the ideal entropy of mixing of ionized and non-ionized monomers. The third term imposes a constraint of fixed pH by introducing a chemical potential coupled to the fraction of charged monomers. The fourth term corresponds to the screened electrostatic interaction energy of the monomers on the polyelectrolyte chain and the dipolar interaction between ion-pairs formed due to counterion condensation. The last term accounts for the charge density fluctuations of the uncondensed counterions. Within the Debye-Huckel mean field theory the latter is given by $\beta F_{fl} = - V \kappa^3/ 12\pi$ \cite{Barrat,McQuarrie}, where $V$ is the solution volume. It is to be noted that $F_{fluc}$ term in Eq. (\ref{free-energy-CI}) is the effective counterion-counterion electrostatic interaction energy contribution to the total free energy within the Debye-Huckel mean field approximation valid in dilute solutions \cite{McQuarrie}. Some of the key steps involved in obtaining the above expressions are summarized in Appendix C. The above equation, when minimized with respect to $f$, results in the following expression:
\begin{eqnarray}
\label{eqn2}
& & N[\log_{10}\frac{f}{1-f} + pH - pK_0] + \frac{f \,l_B \,N^{3/2}\, \Lambda(\kappa_0)}{3\,\pi^2 \,b\, \alpha} + \frac{f^2\, l_B\, N^{3/2}\, \partial_f(\Lambda(\kappa_0))}{6\,\pi^2\, b\, \alpha}+\frac{32\,\sqrt{6\,\pi}\,(1-f)\,l_B^2\,d^4\,N^{1/2}}{27\,b^6\,\alpha ^3} \nonumber \\
&& \hspace{10cm}+ N \left[(2-f)\, \rho \,l_B^3\, \pi\right]^{1/2} = 0.
\end{eqnarray}
Eqs. (\ref{eqn1}) and (\ref{eqn2}) are the key equations of the present work.  In what follows, we provide a numerical solution of these two coupled equations.

\begin{figure}[t]
\centering
\includegraphics[trim=0.2cm 2.0cm 1cm 2.5cm,clip=true,scale=0.9]{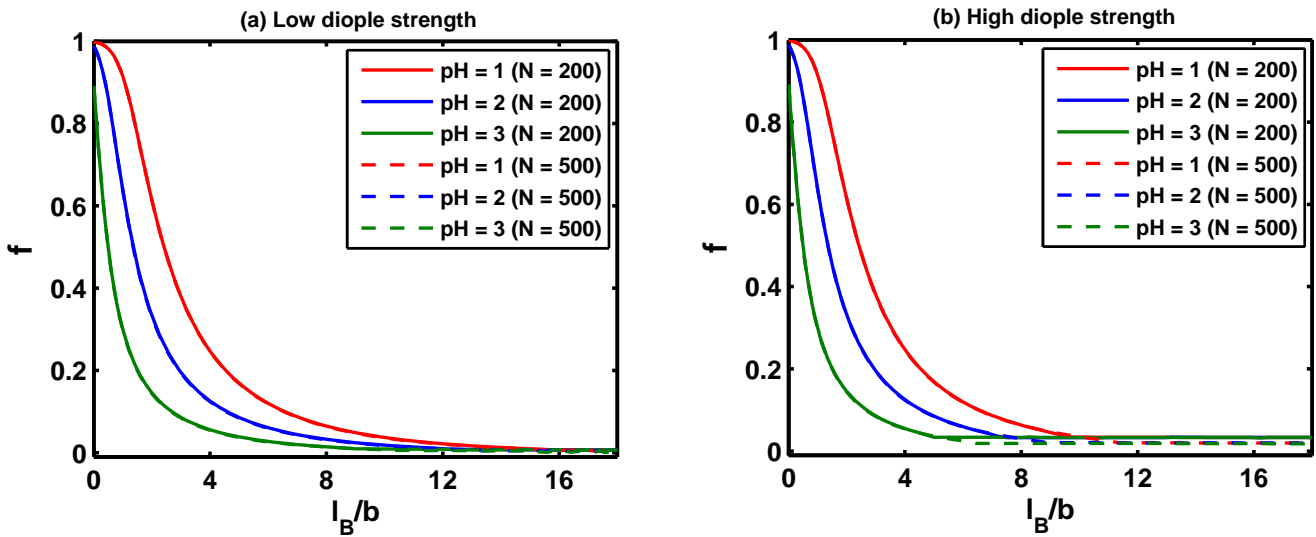}
\caption{The dependence of the solution $pH$ on the variation of the fraction of charges on the polyelectrolyte backbone, $f$, as a function of the increase in the dimensionless strength of the electrostatic interaction characterized by the Bjerrum length, $l_{B}/b$,  for $ N=200$ (solid line) and $500$ (dashed line),  $\rho/b^3=0.001$, $v/b^3 = -0.2$ and $pK_0=4$ and (a) $d/b = 0.01$,  (b) $d/b = 0.5$. As expected, the fraction of charges on the chain backbone are higher at low $pH$ compared to the higher values. The fraction of charges reduce continuously with the increase in the electrostatic strength due to counterion condensation at high electrostatic strength.  }
\end{figure}

\section{Results}

To calculate the fraction of charges on the polyelectrolyte backbone and the size of the weak polyelectrolyte, the non-linear coupled equations, given by Eqs. (\ref{eqn1}) and (\ref{eqn2}), are solved numerically for the dimensionless size ($\alpha$) and  the fraction of charges ($f$) respectively as a function of the dimensionless strength of the electrostatic interaction characterized by the Bjerrum length, $l_{B}/b$. This is done by keeping the dimensionless variables like the number of monomers, $N$, the density of monomers, $\rho/b^3$, the two-body interaction parameter representing the quality of solvent, $v/b^3$,  the dipole interaction strength between two dipoles characterized by the distance between the charge on the monomer and oppositely charged counterions, $d/b$, $pH$ and $pK_0$ constant.

For a weak polyelectrolyte, the fraction of charges, $f$, on the polyelectrolyte backbone are not fixed, but depends on the solution pH \cite{Barrat,Zito,Raphael,Uyaver}. This is depicted in Fig. (1) which shows the variation of the fraction of charges, $f$, along the chain backbone as a function of the dimensionless Bjerrum length, $l_B/b$ at different pH. At $l_B/b = 0$, when the electrostatic interaction between the charged monomers and the counterions is absent and only contribution to the free energy expression in Eq. (\ref{free-energy-CI}) is from the ideal entropy of mixing,  Eq. (\ref{eqn2}) suggests that $pH = pK_0 + \log_{10} \frac{1-f}{f}$. 
As expected, therefore, at very small value of the electrostatic interaction strength, $l_B/b$, the weak polyelectrolyte is fully ionized at very low pH, but show partial ionization at higher pH. With the increase in the Bjerrum length, the electrostatic interaction between oppositely charged counterions and monomers becomes important, as a result of which counterions begin to condense on the chain resulting in the reduction of the effective charge. At large Bjerrum length, all the counterions condense on the chain resulting in the zero effective charge [Fig. (1a)]. Fig. (1) shows that the fraction of charges reduce faster for a solution at higher pH as the chain is only partially charged as compared to very low pH when the chain is fully charged. 

With the increase in the dipole strength, Fig. (1b), there is reduction of the fraction of charges.  However, the effective charge is very close to zero but not exactly zero. The values of the effective charge at large $l_B/b$, for instance, are $f = 0.034$ and  $f = 0.02$ for $N=200$ and $N=500$ respectively in Fig. (1b). A possible reason for this can be that at high dipole strength, $d/b$, the attractive interaction between dipoles [the fourth term in Eq. (\ref{eqn2})] dominates. This results in a compact globular state which is expected to form faster when the dipole strength is high compared to when it is low. At high dipole strength, therefore, the formation of the compact globular state can prevent counterion  condensation to reduce the effective charges to zero.  This also shows that at high electrostatic strength, the coupling of the fraction of charges with the polyelectrolyte size, governs the phenomenon of counterion condensation.

\begin{figure}[t]
\centering
\includegraphics[trim=0.3cm 1.5cm 1cm 2.5cm,clip=true,scale=.9]{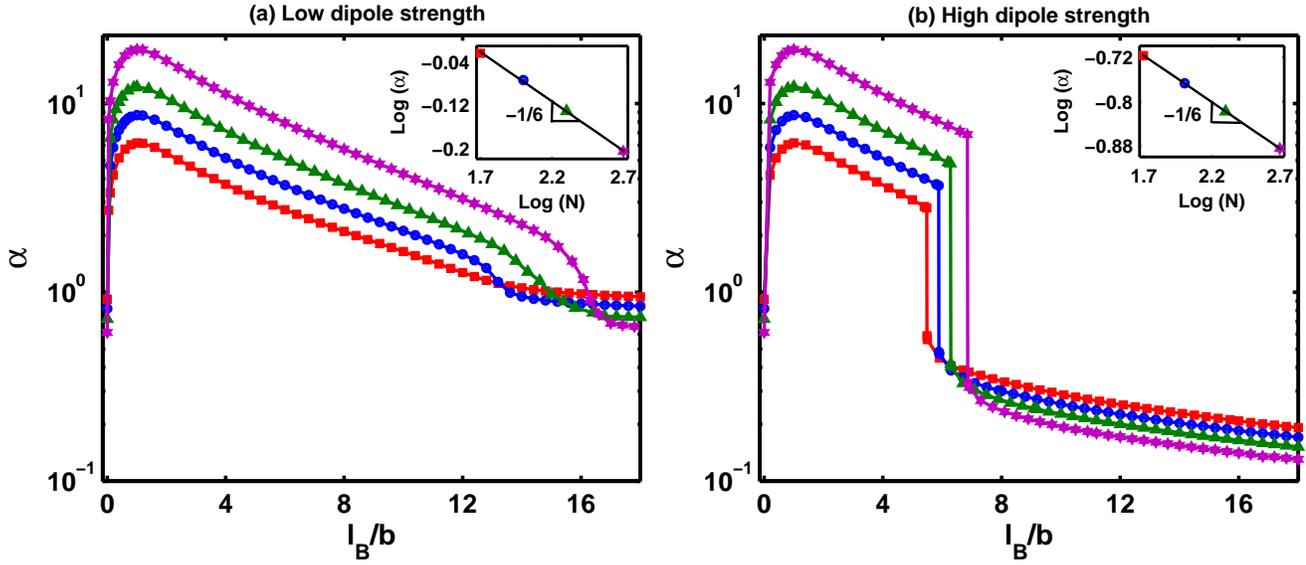}
\caption{ Dependence of the degree of polymerization on the variation of the dimensionless size $ \alpha $ as a function of the dimensionless electrostatic strength, $l_{B}/b$, for $ N=50$ (red), $100$ (blue), $200$ (green) and $500$ (magenta), $\rho/b^3=0.001$, $v/b^3=-0.2$, $pK_0=4$, $pH=1$, and (a)  $d/b=0.01$, (b) $d/b=0.5$. At low dipole strength (a), the reduction in the polyelectrolyte size from the extended state is continuous and the reduced size at high electrostatic strength shows re-entrant poor solvent size. At high dipole strength (b), the collapse transition is discontinuous and the collapsed size is a compact globular state independent of the solvent quality. In the latter cases, a discontinuous transition is observed even for low degree of polymerization. The inlets of (a) and (b) show $\log(\alpha)$ versus $\log(N)$ plots at $l_B/b = 20$. The inlet of (a) shows that $\alpha \varpropto N^{-1/6}$ corresponding to a re-entrant globular state at high $l_B/b$. The inlet of (b) shows that $\alpha \varpropto N^{-1/6}$ corresponding to a collapsed globular state at high $l_B/b$.  Although the slopes in both cases are the same, the difference is in the values of the intercept which is lower in (b) compared to (a) }
\end{figure}

To understand the effects of the degree of polymerization [Fig. (2)], solvent quality [Fig. (3)] and solution pH [Fig. (4)] on the conformational behaviour of weak polyelectrolytes, Figs. (2)-(4) show the variation of the dimensionless size, $\alpha$, as a function of the electrostatic interaction strength, $l_B/b$. A common feature of all three plots is the non-monotonic variation in the polyelectrolyte size with the increase in the electrostatic interaction strength. At $l_B/b =0$, the size of the chain is determined by the quality of solvent, which is a globular [$\alpha < 1$] state corresponding to $\alpha \approx b N^{1/6}/v^{1/3} N^{-1/6}/b^3$ [Fig. (2)]. In this limit, the size is determined by the balance of the third and fourth terms in Eq. (\ref{eqn1}). With the increase in the electrostatic interaction strength $l_B/b$, the polyelectrolyte size swells to a rod-like state [$\alpha \gg 1$] due to electrostatic repulsion between like-charged monomers on the chain, given by the fifth term in Eq. (\ref{eqn1}).  The increase in $N$ leads to much larger increase in the size resulting in a more extended rod-like state [Fig. (2)]. 

When $l_B/b \geq 1$,  counterions begin to condense on the polyelectrolyte backbone resulting in the formation of ion-pairs between charged monomers and oppositely charged counterions, thereby reducing the effective charge of the polyelectrolyte. The reduction in the fraction of charges due to counterion condensation reduces the repulsive strength of the electrostatic interaction [fifth term in Eq. (\ref{eqn1})], as a result of which the size of the polyelectrolyte gradually shrinks. At large Bjerrum length, when the effective charge on the polyelectrolyte is close to zero,  the chain shrinks to its original size. The latter corresponds to  a re-entrant globular state in poor solvents at high $l_B/b$ [Fig. (2a)]. This occurs when the strength of the dipole interaction is low, as a result of which the contribution from the  attractive dipole interaction [sixth term in Eq. (\ref{eqn1})] is insignificant. At low dipole strength, therefore, the reduction in the chain size at large $l_B/b$ is continuous and shows a re-entrant poor solvent state with the size comparable to the original size at $l_B/b =0$ [Fig. (2a)].  The latter is depicted in the inlet of Fig. (2a) which shows that $\alpha \varpropto N^{-1/6}$.

\begin{figure}[t]
\centering
\includegraphics[trim=0.1cm 1.8cm 1cm 2.5cm,clip=true,scale=.9]{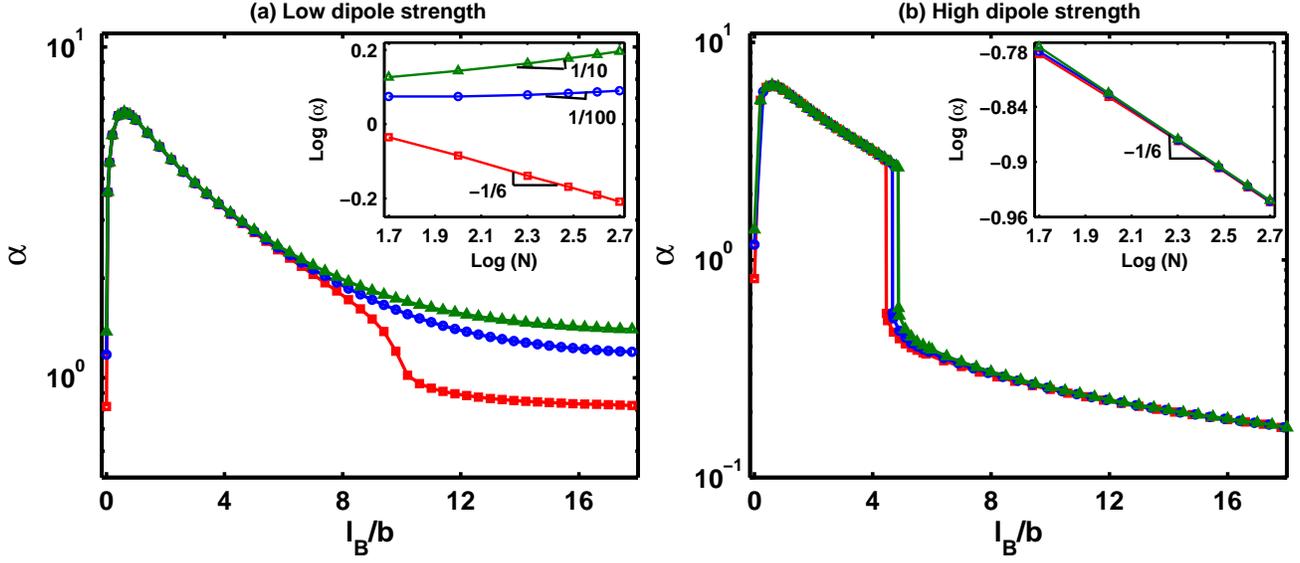}
\caption{ Dependence of the solvent quality on the variation of the dimensionless size $\alpha $ as a function of the dimensionless electrostatic strength, $l_{B}/b$,  for $ N=100$, $\rho/b^3 =0.001$, $v/b^3=-0.2$ (red), $0$ (blue) and $0.2$ (green), $pH=2.5$, $pK_0=4.0$ and (a) $d/b = 0.01$,  (b) $d/b = 0.5$. A non-monotonic conformational behaviour is observed with the increase in the electrostatic strength. At low dipole strength (a), the reduction in the polyelectrolyte size from the extended state is continuous and the reduced size at high electrostatic strength shows re-entrant poor (red), theta (blue) and good (green) solvent size. At high dipole strength (b), the collapse transition is discontinuous and the collapsed size is a compact globular state independent of the solvent quality. The inlets of (a) and (b) show $\log(\alpha)$ versus $\log(N)$ plots at $l_B/b = 20$ for poor (red), theta (blue) and good (green) solvents . The inlet of (a) shows that $\alpha \varpropto N^{-1/6}$, $\alpha \varpropto N^{-1/100} \approx N^0$ and $\alpha \varpropto N^{1/10}$ corresponding to a re-entrant globular, ideal coil and coiled state at high $l_B/b$ respectively. The inlet of (b) shows that $\alpha \varpropto N^{-1/6}$ corresponding to a collapsed globular state at high $l_B/b$ independent of solvent quality. }
\end{figure}

At high dipole strength [Fig. (2b)],  counterion condensation sets in at $l_B/b \geq 1$,  which results in the reduction of the fraction of charges on the polyelectrolyte backbone. As a result of this, the repulsive strength of the electrostatic interaction [fifth term in Eq. (\ref{eqn1})] reduces resulting in the shrinkage of the polyelectrolyte size. The initial shrinkage in the polyelectrolyte size occurs due to charge renormalization because of  counterion condensation. With the further increase in $l_B/b$, the attractive interaction between dipoles [sixth term in Eq. (\ref{eqn1})] overcomes the reduced strength of repulsive  electrostatic interaction [fifth term in Eq. (\ref{eqn1})], resulting in an abrupt shrinkage of the polyelectrolyte size to a compact globular state. At high dipole strength, therefore, the chain collapse to compact globular state is discontinuous [Fig. (2b)], and occurs due to attractive dipole interactions between ion-pairs formed due to condensed counterions.  The scaling of the polyelectrolyte size with respect to $N$ at high $l_B/b$ is depicted in the inlet of Fig. (2b) which shows that $\alpha \varpropto N^{-1/6}$ corresponding to a compact globular state. The fact that the compact globular size is less than the re-entrant globular size is captured by the intercept of $\log(\alpha)$ versus $\log(N)$ plot [inlet of Fig. (2b)] which is less compared to the re-entrant globular state [inlet of Fig. (2a)]. Interestingly, Fig. (2b) shows that at high dipole strength the nature of conformational transition at high $l_B/b$ is discontinuous even for low $N$. This point is discussed in more detail later. 


Fig. (3) shows the non-monotonic variation of the dimensionless size $\alpha$ as a function of the solvent quality. As before, at $l_B/b =0$ the size of the chain is determined by the quality of solvent, which is in coiled [$\alpha >1$], ideal coiled [$\alpha \approx 1$] or globular [$\alpha < 1$] state corresponding to good [$v >0$], theta [$v=0$] or poor [$v<0$] solvents respectively. In this limit, the size is determined by the balance of the first four terms in Eq. (\ref{eqn1}) which yields $\alpha \varpropto N^{1/10}$, $\alpha \varpropto N^0$ and $\alpha \varpropto N^{-1/6}$ for good, theta and poor solvents respectively. With the increase in the electrostatic interaction strength $l_B/b$, irrespective of the quality of solvent, the polyelectrolyte size swells to a rod-like state [$\alpha \gg 1$] due to electrostatic repulsion between like-charged monomers on the chain, given by the fifth term in Eq. (\ref{eqn1}). The size in this limit is governed by the first five terms in  Eq. (\ref{eqn1}). At large Bjerrum length, when the effective charge on the polyelectrolyte is close to zero due to counterion condensation,  the chain shrinks to its original size. The latter corresponds to  a re-entrant coiled state in good solvents, ideal coiled state in theta solvents and globular state in poor solvents, the scaling of which  is shown in the inlet of Fig. (3a) for high $l_B/b$ .  Thus, at low dipole strength the reduction in the chain size at high $l_B/b$ is continuous and, depending on the solvent quality, there is a re-entrant good, theta or poor solvent state [Fig. (3a)]. At high dipole strength [Figs. (3b)],  counterion condensation results in a discontinuous collapse transition from a rod-like state to a compact globular state. Thus, irrespective of the quality  of solvent, the collapsed state is a compact globular state of size less than the original size [Fig. (3b)]. This is shown in the inlet of Fig. (3b) which yields the same scaling for the compact globular size, $\alpha \varpropto N^{-6}$, independent of the solvent quality.

\begin{figure}[t]
\center
\includegraphics[trim=0.1cm 1.6cm 1cm 2.5cm,clip=true,scale=.9]{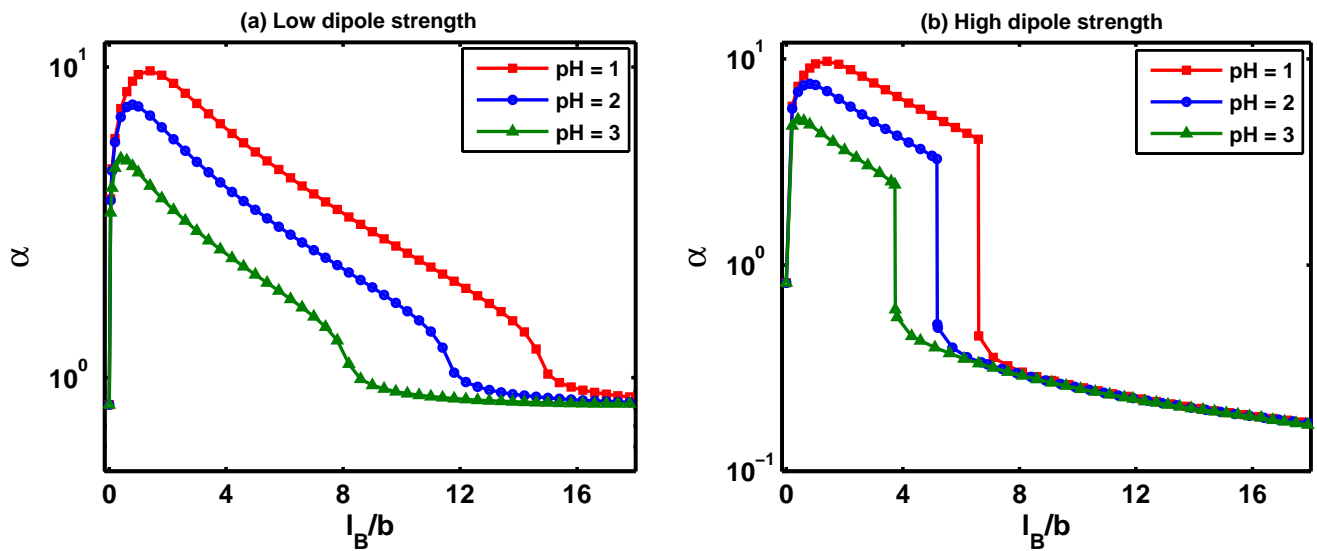}
\caption{Dependence of the solution pH on the variation of the dimensionless size $ \alpha $ as a function of the dimensionless electrostatic strength, $l_{B}/b$, for $ N=100 $, $\rho/b^3=0.001$, $v/b^3=-0.2 $, $pH=1, 2, 3$, $pK_0=4$ and (a) $d/b=0.01$, (b) $d/b=0.5$. The polyelectrolyte chain extension is higher at low pH because of the higher fraction of charges on the chain backbone. At low dipole strength (a), the reduction in the polyelectrolyte size from the extended state is continuous and the reduced size at high electrostatic strength shows re-entrant poor solvent size. At high dipole strength (b), the collapse transition is discontinuous and the collapsed size is a compact globular state independent of the solvent quality.  }
\end{figure}

Fig. (4) shows the non-monotonic variation of the dimensionless size as a function of the dimensionless Bjerrum length in a poor solvent at different $pH$. With the decrease in the solution pH, the polyelectrolyte chain becomes more ionized resulting in a more extended rod-like state [Fig. (4)]. As before, when $l_B/b \geq 1$,  counterion condensation occurs resulting in the reduced effective charge of the polyelectrolyte chain. At low dipole strength, this results in a continuous reduction in the polyelectrolyte size to a re-entrant poor solvent state at high electrostatic strength [Fig. (4a)] due to charge renormalization. At high dipole strength, the attractive interaction between dipoles results in a discontinuous  chain collapse to a size which is smaller than the original size at $l_B/b =0$ [Fig. (4b)].

\section{Summary and Conclusion}

In this work we have presented a self-consistent  variational theory for the conformational behaviour of weak polyelectrolytes which includes, crucially, the variation in dipole strength of the ion-pairs formed due to counterion condensation. The conformational behaviour is studied as a function of the solvent quality, the solution pH or the degree of polymerization and captures several qualitative features of viscometric and light scattering measurements on  weak polyelectrolytes \cite{Klooster}. This include non-monotonic variation of the polyelectrolyte size as a function of the increase in the electrostatic strength; a continuous reduction in the polyelectrolyte size at low dipole strength   corresponding to smaller and less polarizable condensed counterions; a discontinuous collapse transition at high dipole strength, corresponding to larger and more polarizable condensed counterions;  a re-entrant good, theta and poor solvent size at low dipole strength due to charge renormalization in the presence of counterion condensation at high  electrostatic strength; a discontinuous collapse transition to compact globular state, independent of the solvent quality, at high dipole strength due to attractive dipole interaction between ion-pairs. Here, as in experiments, the discontinuous chain collapse at higher dipole strength, which abruptly reduce the size of a weak polyelectrolyte, occurs at much lower value of the electrostatic interaction strength compared to the case when the dipole strength is low.  

The self-consistent mean field approach of the uniform expansion method presented here has been used earlier to study the effects of counterion condensation in strong polyelectrolytes \cite{Muthukumar}. However, the main difference between the present work  and the previous work in Ref. \cite{Muthukumar} is that in the latter work the effect of counterion condensation  is not due to the dipolar interactions between condensed counterions, but due to disparity between the bulk dielectric constant and local dielectric constant close to the chain backbone. The neglect of dipolar interactions results in a continuous reduction in the chain size  as a function of the Bjerrum length. This can not explain the experimental results which observe a shift from the continuous to the discontinuous transition with the increase in the size and polarizability of the counterions.

Thus, a continuous reduction in the polyelectrolyte size and a discontinuous collapse to a compact globular state at high electrostatic strength, observed in independent simulations in references \cite{Limbach,Sunil} and \cite{Anup} respectively, are subsumed and unified in the present theory which shows a shift from continuous to discontinuous collapse transition with the increase in the dipole strength, in agreement with experiments \cite{Klooster}.  The present work shows that the dipole strength of the ion-pairs, formed due to condensed counterions at high electrostatic strength, plays a significant role in determining the nature of collapse transition and the size of the collapsed state at high electrostatic strength. It also shows that even for low degree of polymerization, the attractive interaction between ion-pairs of  higher dipole strength can induce discontinuous collapse transition. Thus, the reason for either observing continuous collapse transition at low degree of polymerization \cite{Sunil} or discontinuous collapse transition at high degree of polymerization \cite{Anup} in recent simulations can not be merely attributed to finite size effects, but has to considered in relation to the dipole strength of the ion-pairs  formed between charged monomers and oppositely charged condensed counterions at high electrostatic strength.

\appendix
\section{Derivation of the two-body interaction strength in the presence of  the dipolar interactions}
\renewcommand{\theequation}{\thesection\arabic{equation}}
\vskip 0.05in

The interaction potential in the presence of counterions is 
\begin{equation}
U(r) = U_{rep}(r)+U_{att}(r)+U_{d}(r)
\end{equation}
where, $U_{rep}(r)$ and $U_{att}(r)$ are the hardcore and weak attractive potentials respectively. $U_{d}(r)$ is the interaction energy between two dipoles separated at a distance $r$, which is assumed to be weak. 
The Mayer $f$-function is given by
\begin{equation}
f(r) = e^{-\beta \lbrace U_{rep}(r)+U_{att}(r)+U_{d}(r)\rbrace} - 1.
\end{equation}
In terms of the Mayers $f$-function, the expression for the two-body interaction strength is given by
\begin{eqnarray}
v_2 & = & - 4 \pi \int_0^{\infty} d{r} r^2 f(r) \nonumber \\ 
&=& -4 \pi \int_0^b dr r^2 f(r) - 4 \pi \int_b^{\infty} dr r^2 f(r).
\end{eqnarray}
For $r < b$,\, $\beta U_{rep}(r) \gg 1$, only the repulsive potential dominates resulting $f(r) = e^{-\beta U_{rep}(r)} - 1 \cong -1$ . For $r > b$, on the other hand, the attractive potentials dominate and $\beta (U_{att}(r)+U_{d}(r)) < 1$, which results in $f(r) = e^{-\beta \lbrace U_{att}(r)+U_{d}(r)\rbrace} - 1 \cong  -\beta \lbrace U_{att}(r)+U_{d}(r)\rbrace $.
Therefore, the modified expression for the two-body interaction term  in the presence of dipolar interactions can be approximately written as
\begin{eqnarray}
v_2 &\approx & 4\pi\int _{0}^{b} dr \,r^2 + \frac{4\pi}{k_B T}\int _{b}^{\infty} dr \,r^2\,U_{att}(r) + \frac{4\pi}{k_B T}\int _{b}^{\infty} dr \,r^2\,U_{d}(r) \nonumber\\[1.5ex] 
&\approx & \left(1-\frac{\theta}{T}\right) b^3 + \frac{4\pi}{k_B T}\int _{b}^{\infty} dr \,r^2\, U_{d}(r).  
\end{eqnarray}
where $\displaystyle{\theta \approx - (b^3\,k_B)^{-1}\int _{b}^{\infty} dr \,r^2 \, U_{att}(r)}$ is called the $\theta$-temperature. The term $\displaystyle  \left(1-\frac{\theta}{T}\right) b^3$ is the excluded volume, $v$, and it can either be positive, negative or zero depending on whether $T > \theta$,  $T < \theta$ or $T= \theta$ respectively. Using the expression for $\beta U_{d}(r) = - 4 \pi (1-f)^2\, l_B^2 \,d^4/3\, r^6$, $v_2$ can be written as 
\begin{eqnarray}
\label{modv}
v_2 & \approx & v+4\pi\int _{b}^{\infty} dr \,r^2\,\left(-\frac{4\,\pi\,(1-f)^2\,l_B^2\,d^4}{3\,r^6}\right) \nonumber\\ \nonumber\\
&=& v-\frac{16\,\pi ^2\,(1-f)^2\,l_B^2\,d^4}{9\,b^3}. 
\end{eqnarray}
\section{Uniform Expansion Method for weak polyelectrolytes in the presence of dipolar interactions}
\renewcommand{\theequation}{\thesection\arabic{equation}}
\vskip 0.05in

The detailed calculations of the entropic term are fairly standard and can be found in Refs. \cite{Doi} and \cite{Singh}. The corresponding result is 
\begin{equation}
\label{entropic}
\left<{\bf R}^2 \right> \left< {\cal H}_0 - {\cal H}_1 \right> - \left<{\bf R}^2\left({\cal H}_0-{\cal H}_1\right)\right>=Nb_1^4\left(\frac{1}{b_1^2} - \frac{1}{b^2}\right).
\end{equation} \\ \paragraph*{}
To calculate  $\left<{\bf R}^2\right> \left<{\cal H}_2\right> - \left<{\bf R}^2\left({\cal H}_2\right)\right>$,  we use
\begin{equation}
\delta\left[{\bf R}_n-{\bf R}_m\right]= \int_{-\infty}^{\infty}\frac{d^3{\bf k}}{(2\pi)^3} \,e^{i{\bf k}\cdot|{\bf R}_n-{\bf R}_m|}
\end{equation}
and the result
\begin{equation}
\left<e^{i{\bf k}\cdot|{\bf R}_n-{\bf R}_m|}\right>=e^{-{\bf k}^2(n-m)b_1^2/6}.
\end{equation}  
This leads to the following result:
\begin{equation}
\label{2body}
\left<{\bf R}^2\right> \left<{\cal H}_2\right> - \left<{\bf R}^2\left({\cal H}_2\right)\right> = \frac{v_2\, b_1^4}{12{\pi}^2}\int_{0}^{N}dn\int_{0}^{N}dm\int_{0}^{\infty}dk\, k^4(n-m)^2 \,e^{-k^2(n-m)b_1^2/6}
\end{equation}
where
\begin{equation}
v_{2}=v-\frac{16\,\pi ^2\,(1-f)^2\,l_B^2\,d^4}{9\,b^3}.
\end{equation}
$v$ is the strength of the excluded volume. \\[1ex]

To calculate $\left<{\bf R}^2\right> \left< {\cal H}_c\right>  - \left<{\bf R}^2 {\cal H}_c\right>$,  we use the following identity:

\begin{equation}
\label{eqn11}
\frac{\exp({-{\kappa}|{\bf R}_n - {\bf R}_m|})}{|{\bf R}_n - {\bf R}_m|} = \int_{-\infty}^{\infty}\frac{d^3{\bf k}}{(2\pi)^3}\,\frac{\exp({i{\bf k}\cdot|{\bf R}_n - {\bf R}_m|})}{{\kappa}^2+{\bf k}^2}.
\end{equation}
Further simplification results in 
\begin{equation}
\left<{\bf R}^2\right> \left< {\cal H}_c\right>  - \left<{\bf R}^2 {\cal H}_c\right>  = \frac{f^2 b_1^4 l_B}{6{\pi}^2}\int_{0}^{N} dn \int_{0}^{n} dm \int_{0}^{\infty} \frac{k^4 (n-m)^2}{({\kappa}^2+k^2)} \,e^{-k^2 (n-m)b_1^2/6}.
\end{equation}
Defining the dimensionless quantities $k_0=k\,N^{1/2}\,b_1$, ${\kappa}_0=\kappa \, N^{1/2}\, b_1$, $m=nx$ and $n=Ny$ the above equation can be written in the dimensionless form
\begin{equation}
\label{screencoulombic}
\left<{\bf R}^2\right> \left< {\cal H}_c\right>  - \left<{\bf R}^2 {\cal H}_c\right>  = \frac{f^2\,l_B\, N^{5/2}\,b_1}{6{\pi}^2}\int_{0}^{1} dy \int_{0}^{1} dx \int_{0}^{\infty}dk_0\frac{k_0^4\, (1-x)^2 \,y^3}{({\kappa}_0^2+k_0^2)} \,e^{-k_0^2\,(1-x)\,y/6}.
\end{equation}     \\ \paragraph*{}
The uniform expansion method of Edwards and Singh \cite{Singh} does not include the three-body interaction term. A detailed calculation, which can be found in Ref. \cite{Dua}, yields
\begin{equation}
\left< {\bf R}^2 {\cal H}_3 \right> = \frac{w}{6} \int_{0}^{N} dn \int_{0}^{N} dm \int_{0}^{N} dl \left< R^2\,\delta[{\bf R}_n - {\bf R}_m]\, \delta[{\bf R}_m - {\bf R}_l] \right>.
\end{equation}~

In terms of the wave vectors ${\bf k}$ and ${\bf q}$, the above equation can be rewritten as

\begin{equation}
\frac{w}{6} \int_{-\infty}^{\infty} \frac{d^3{\bf k} }{(2\pi)^3}\int_{-\infty}^{\infty} \frac{d^3{\bf  q}}{(2\pi)^3}\int_{0}^{N} dn \int_{0}^{N} dm \int_{0}^{N} dl \left< {\bf R}^2\, e^{i {\bf k} \cdot |{\bf R}_n - {\bf R}_m|}\, e^{i {\bf q} \cdot |{\bf R}_m - {\bf R}_l|} \right>.
\end{equation}~

Expanding the mean square end-to-end distance and the probability distribution in terms of internal coordinates and using the fact that the probability distribution is Gaussian, we get

\begin{eqnarray}\label{3body1}
\left< {\bf R}^2 {\cal H}_3 \right> &=& \frac{w}{6} \int_{-\infty}^{\infty} \frac{d^3{\bf k} }{(2\pi)^3}\int_{-\infty}^{\infty} \frac{d^3{\bf  q}}{(2\pi)^3}\int_{0}^{N} dn \int_{0}^{N} dm \int_{0}^{N} dl\,\left[Nb_1^2  - \frac{{\bf k}^2\,\left(n-m\right)^2\,b_1^4}{3} - \frac{{\bf q}^2\,\left(m-l\right)^2\,b_1^4}{3} \right] \nonumber\\ \nonumber\\
&& \hspace{5cm} e^{-{\bf k}^2 (n-m)b_1^2/6}\, e^{-{\bf q}^2 (m-l)b_1^2/6}.
\end{eqnarray} ~

Repetition of the above procedure to calculate $\left< {\bf R}^2 \right> \left< {\cal H}_{3} \right>$ gives 

\begin{equation}\label{3body2}
\left< {\bf R}^2 \right> \left< {\cal H}_{3} \right> = N b_1^2\,\frac{w}{6} \int_{-\infty}^{\infty} \frac{d^3{\bf k} }{(2\pi)^3}\int_{-\infty}^{\infty} \frac{d^3{\bf  q}}{(2\pi)^3}\int_{0}^{N} dn \int_{0}^{N} dm \int_{0}^{N} dl\,\,  e^{-{\bf k}^2 (n-m)b_1^2/6}\, e^{-{\bf q}^2 (m-l)b_1^2/6}.
\end{equation}~

Subtracting Eq. (\ref{3body1}) form Eq. (\ref{3body2}) one obtains

\begin{eqnarray}\label{3body}
\left< {\bf R}^2 \right> \left< {\cal H}_{3} \right> - \left< {\bf R}^2 {\cal H}_3 \right> &=&  \frac{w}{6} \int_{-\infty}^{\infty} \frac{d^3{\bf k} }{(2\pi)^3}\int_{-\infty}^{\infty} \frac{d^3{\bf  q}}{(2\pi)^3}\int_{0}^{N} dn \int_{0}^{N} dm \int_{0}^{N} dl\,\left[\frac{{\bf k}^2\,\left(n-m\right)^2\,b_1^4}{3} + \frac{{\bf q}^2\,\left(m-l\right)^2\,b_1^4}{3} \right] \nonumber\\ 
&& \hspace{7.5cm} e^{-{\bf k}^2 (n-m)b_1^2/6}\, e^{-{\bf q}^2 (m-l)b_1^2/6}  \nonumber\\ \nonumber\\
&=&\frac{w \,b_1^4}{12 \pi^4} \int_{0}^{\infty}dk \int_{0}^{\infty} dq \int_{0}^{N} dn \int_{0}^{n} dm \int_{0}^{m} dl  \left[(n - m)^2 k^4 q^2 + (m - l)^2 k^2 q^4\right]\nonumber\\
& &  \hspace{7.5cm} e^{-k^2\,(n - m)\,b_1^2/6} e^{-q^2\,(m - l)\,b_1^2/6}.
\end{eqnarray}

The last integral diverges as $q \rightarrow \infty$.  The divergence can be removed by introducing an upper cut-off for the wave number $q$, the details of which are discussed in Ref. \cite{Dua}. The integrations in Eqs. (\ref{entropic}), (\ref{2body}), (\ref{screencoulombic}) and (\ref{3body}) can easily be carried out. Substitution of the results in Eq. (\ref{Hamiltonian}) gives the following variational equation.

\begin{equation}
\label{eqn19}
N{b_1}^2\left(1-\frac{{b_1}^2}{b^2}\right) + \frac{\sqrt{6}\,\left(v-\frac{16\,\pi ^2\,(1-f)^2\,l_B^2\,d^4}{9\,b^3}\right)\,N^{3/2}}{ \pi^{3/2} {b_1}} + \frac{f^2 l_B N^{5/2} b_{1}\Lambda(\kappa_0)}{6\pi^2} + \frac{w N}{{b_{1}}^4} = 0,
\end{equation}

where 

\begin{equation}
\Lambda(\kappa_0)=\int_{0}^{\infty}dk_0 \int_{0}^{1} dy \int_{0}^{1} dx \frac{k_0^4\, (1-x)^2 \,y^3}{({\kappa}_0^2+k_0^2)} \,e^{-k_0^2\,(1-x)\,y/6}.
\end{equation}~

Dividing the above equation by $Nb_{1}^2$, taking $\alpha=b_{1}/b$ and considering $w\sim b^6$, we obtain the variational equation given by Eq. (\ref{eqn1}). \\ \paragraph*{}
The variational equation in the Edwards-Singh method directly estimates $\frac{\partial F}{\partial R}$ \cite{Prasanta}. Therefore the effective free energy of the polyelectrolyte chain can be estimated employing the method of thermodynamic integration which leads to Eq. (\ref{free-energy1}).

\section{ Calculation of the ideal entropy of mixing}
\renewcommand{\theequation}{\thesection\arabic{equation}}
\vskip 0.1in

If $ N $ is the degree of polymerization and $ M $ is number of ionized monomers, then the degree of ionization is $ f=\frac{M}{N} $. For a weak polyelectrolyte at a given pH, the ionized and non-ionized monomers are randomly distributed along the chain backbone. At vanishingly small $l_B$, therefore, the ideal entropy of mixing of the ionized and non-ionized monomers can be written as 

\begin{eqnarray}
S_{mix} &=& k_B\,\log_{10} \left[\frac{N!}{\left(N-M\right)!\,M!}\right]. 
\end{eqnarray}
Using Stirling's approximation for large $N$ and $M$,
\begin{eqnarray}
S_{mix} &=& k_B\,\left[-N\,\log_{10} \frac{\left(N-M\right)}{N}+M\,\log_{10} \left(\frac{N-M}{M}\right)\right]\nonumber \\ \nonumber \\
&=& -k_B\,N\left[f\,\log_{10} f+\left(1-f\right)\,\log_{10} \left(1-f\right)\right].
\end{eqnarray} 

The free energy of mixing is given by
\begin{eqnarray}
F_{mix} &=& -T\,S_{mix} \nonumber \\ \nonumber \\  
&=&  N\,k_B\,T\,\left[f\,\log_{10} f+\left(1-f\right)\,\log_{10} \left(1-f\right)\right].
\end{eqnarray}
Therefore, 
\begin{equation}
\beta F_{mix}= N\,\left[f\,\log_{10} f+\left(1-f\right)\,\log_{10} \left(1-f\right)\right].
\end{equation}

The free energy due to charge density fluctuations is given by
\begin{eqnarray}
\beta\,F_{fl} &=& -\frac{\kappa ^3}{12\,\pi}V \nonumber \\ \nonumber \\ 
&=& -\frac{\left(4\,\pi\right)^{3/2}V}{12\,\pi}\,l_B^{3/2}\,\rho ^{3/2}\,\left[f+2\,\left(1-f\right)\right]^{3/2} \nonumber \\ \nonumber \\
&=& -\frac{\left(4\,\pi\right)^{1/2}V}{3}\,l_B^{3/2}\,\rho ^{3/2}\,\left(2-f\right)^{3/2}.
\end{eqnarray}

Differentiating with respect to $ f $ gives,

\begin{equation}
\partial_f \left(\beta\,F_{fl}\right)= N\,\pi ^{1/2}\,l_B^{3/2}\,\left[\rho\,\left(2-f\right)\right]^{1/2},
\end{equation}
where $ N=\rho\,V $.

\begin{acknowledgments}
PK acknowledges the financial support from the University Grants Commision (UGC), Government of India.
\end{acknowledgments}

\end{document}